\documentclass[aps,showpacs,showkeys,superscriptaddress,twocolumn]{revtex4}%
\newcommand{\be}{\begin{equation}}
\newcommand{\ee}{\end{equation}}
\newcommand{\bea}{\begin{eqnarray}}
\newcommand{\eea}{\end{eqnarray}}

\date{\today}
\begin{document}
\vspace{1.0cm}
\preprint{NT@UW-09-15}

\vspace{.50cm}

\title{Nucleon-nucleon charge symmetry breaking and the 
$dd\rightarrow\alpha\pi^0$ reaction}
\author{A. C. Fonseca}
\affiliation{Centro Fisica Nuclear, Universidade de Lisboa, 1649-003 Lisboa, Portugal $\quad$}
\author{R. Machleidt}
\affiliation{Department of Physics, University of Idaho, Moscow, Idaho 83844, USA
 $\quad$ }
\author{G. A. Miller}
\affiliation{Department of Physics, University of Washington, Seattle,
WA 98195-1560, USA}

\begin{abstract}
We show that using parameters consistent with the  charge symmetry violating difference between the strong $nn$ and $pp$ scattering lengths provides significant constraints on the amplitude for the   $dd\rightarrow\alpha\pi^0$ reaction.
\end{abstract}
\maketitle

 The concepts of charge independence and charge symmetry
 provide powerful tools in organizing the multiplet 
 structure of  hadrons and nuclei. These 
symmetries are not perfect; diverse small but interesting violations 
have been discovered  
\cite{Miller:1990iz,mosreview}.  
Our concern
here is with the breaking of charge symmetry (CS). 
This symmetry is defined as invariance under a
 rotation by $180^\circ$ around the
2-axis in isospin space.
In quantum chromodynamics (QCD), CS implies  that  dynamics are invariant 
under the exchange of the up and down quarks~\cite{Miller:1990iz}.
However, since the up and down quarks do have different masses
($m_u\neq m_d$)~\cite{mumd}, the QCD Lagrangian is not charge symmetric.
This symmetry violation is called charge symmetry breaking (CSB).
The different electromagnetic interactions of the up and down quarks also cause CSB as well as the breaking
of charge independence.
Observing the effects of CSB interactions therefore provides
a probe of $m_u$ and $m_d$, once the electromagnetic interactions are treated.

\vspace{0.1cm} 
It has long been known that  CSB is violated in the $^1S_0$ state of nucleon-nucleon scattering, 
with $a_{pp}-a_{nn}\equiv\Delta a=1.5\pm 0.5$ fm~\cite{Miller:1990iz},
where $a$ denotes the scattering length.
The $nn$ interaction is more attractive than the $pp$. There are a variety of
explanations for this using meson exchange mechanisms~\cite{LM98,MM01}. 
Each of these mechanisms involving the strong interaction can be traced to the mass
 difference between the up and down quarks. Nucleon-nucleon potentials that are consistent
 with this scattering length difference
 are successful in reproducing (along with electromagnetic effects)
 the measured binding energy differences between mirror nuclei~\cite{Bra88,MM01}.
It is interesting to search for further manifestations of the up-down quark mass difference.

Two exciting observations of CSB in experiments
involving the production of neutral pions have stirred  interest in this subject.  CSB  was observed
in the reaction $np\to d\pi^0$ at
TRIUMF by measuring 
the CSB forward-backward
asymmetry of the differential cross section as 
$A_{\rm fb}=[17.2\pm8({\rm stat})\pm5.5({\rm sys})]\times 10^{-4}$
~\cite{Opper:2003sb}. Furthermore,
 the final experiment at the IUCF Cooler ring reported
a  relatively large  $dd\to\alpha\pi^0$ cross section 
($\sigma=12.7\pm2.2$~pb at $T_d=228.5$~MeV and $15.1\pm3.1$~pb at 231.8~MeV)
\cite{Stephenson:2003dv}.
The $dd\to\alpha\pi^0$ reaction 
violates CS since the deuterons and the $\alpha$-particle
are self-conjugate under the CS operator, with a positive
eigenvalue, while the neutral pion wave function changes sign.

The study of CSB $\pi^0$  production reactions
presents an exciting new opportunity to  determine the influence of 
quark masses in nuclear physics, and to use  effective
field theory (EFT) to improve 
 understanding of 
how QCD works~\cite{mosreview}.
This is because chiral symmetry of QCD determines the form of pionic interactions.
 Electromagnetic CSB  is typically of the same
order of magnitude as the strong one,  and also can be handled using EFT.
The EFT for the Standard Model at momenta comparable to the pion mass, $Q\sim
m_\pi$, is chiral perturbation theory ($\chi$PT)~\cite{ulfreview}.
This EFT has been extended to
pion production~\cite{cpt0,Gardestig,Lensky:2005jc,hanhartreview,Gardestig:2005sn},
where typical momenta are $Q\sim \sqrt{m_\pi M}$, with $M$ the nucleon mass.
 EFT using the operators of \cite{vkiv}
was used to correctly predict the sign of
the forward-backward asymmetry in $np\to d\pi^0$ \cite{vKNM}. 

The purpose of the present note is to use information regarding CSB in the nucleon-nucleon system
to constrain or inform the calculations of the 
$dd\rightarrow\alpha\pi^0$ reaction. 
We begin by 
describing the pion production calculations and then show how the nucleon-nucleon CSB is 
relevant for this calculation.
We constrain the parameters by the requirement that NN CSB is consistent with observation.
It turns out that this constraint reduces considerably the uncertainty in the predictions for the
$dd\rightarrow\alpha\pi^0$ reaction.

\vspace{0.1cm} 
Attempts to understand
 the  $dd\to\alpha\pi^0$ reaction began with a survey
of various mechanism using  initial state plane wave functions 
and simplified final state wave functions \cite{Gardestig}.
Next, 
 recent significant
advances in  four-body theory \cite{nogga,Fonseca99}
were  used to
include the effects of  deuteron-deuteron interactions in the initial state,
and to use bound-state wave functions with realistic two- and three-nucleon
interactions \cite{csbnogga}. 
The resulting calculations  are hybrid: the CSB pion-production operators
are constructed using EFT, but the strong interactions are not. The result was
that a cross section of the 
experimentally measured size could be obtained using leading order (LO) 
and  next-to-next-to LO (NNLO) pion production operators. 
However,  not all NNLO diagrams were included. 
A complete analysis would require a careful treatment of  loop diagrams.

\begin{figure}[t] 
\vspace{2.5cm}
\includegraphics{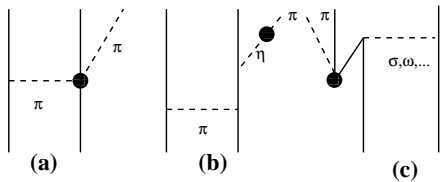}
\caption[]{Diagrams of  $np\to d\pi^0$; the solid circle indicates CSB.}
\label{ddlettdiag}
\end{figure}

The calculation made use of a variety of CSB mechanism
which we will briefly review now.
At formally leading order, there is only one contribution,
represented by Fig.~\ref{ddlettdiag}a:
pion rescattering in which the
CSB occurs through the seagull pion-nucleon terms
linked to the nucleon-mass splitting.

There is no next-to-leading order (NLO) contribution. 
At NNLO,
suppressed by ${\cal O}(m_\pi/M)$, there exists a recoil correction of the LO term, labelled by
${\cal M}_{\rm rec}$. This is determined by the same parameters as the LO term.
 At the same order there are other operators.
A term arises in which
a one-body CSB operator ($\propto \beta_1+\bar{\beta}_3)$
is sandwiched between initial and final state wave functions, as illustrated
in, {\it e.g.}, Fig.~\ref{ddlettdiag}b. We refer to this  as the one-body term (${\cal M}_{1b}$).
The terms $\beta_1 = O(\epsilon m_\pi^2/M^2)$ and
$\bar\beta_3 = O(\alpha/\pi)$ arise from, respectively,
the quark-mass-difference and
electromagnetic contributions to the isospin-violating pion-nucleon coupling.
Neither $\beta_1$ nor $\bar\beta_3$ can be extracted from experiment
yet. These terms were estimated   by modeling~\cite{vKFG} $\beta_1$ 
by $\pi$-$\eta$ mixing, see Fig.~\ref{ddlettdiag}b,
\begin{equation}
\beta_1=\bar{g}_\eta\langle\pi^0|H|\eta\rangle/m_\eta^2, \label{beta1}
\end{equation}
where $\langle\pi^0|H|\eta\rangle=-4200$~MeV$^2$
is the $\pi$-$\eta$--mixing matrix
element~\cite{mesmix}, and
$\bar{g}_\eta=g_{\eta NN} f_\pi/M$
the $\eta$-nucleon coupling constant.
Nucleon-nucleon elastic scattering 
 data show  little sensitivity to $\eta$ exchange and
high-accuracy fits
can be achieved \cite{Machleidt:2000ge} using  $g_{\eta NN}^2/4\pi=0$.
Indeed, the
possibility  of a vanishing coupling constant had been  raised earlier by 
the detailed analysis of $NN$ total cross sections and
$p\bar{p}$ data using dispersion relations \cite{Grein:1979nw} resulting in
 $g_{\eta NN}^2/4\pi \approx 0$.
Also, in the Bonn full model~\cite{MHE87}
it is found that
 $g_{\eta NN}^2/4\pi \approx 0$ 
is consistent with the NN scattering data.
A value of  $g_{\eta NN}^2/4\pi = 0.51$  was used in Ref.~\cite{csbnogga}.
\vspace{0.1cm}

\vspace{0.1cm} 
The effects of electromagnetic 
interactions as well as strong CSB were included in computing the
 $\alpha$ particle wave functions.
These  interactions  
generate a small
isospin $T=1$ component of the wave function that  enables
a non-zero contribution of 
charge-symmetry conserving (CSC) production operators. The one-body
operator was used to generate $\pi^0$ production \cite{csbnogga}.

A number of other CSB mechanisms enter at N$^3$LO or higher,
including loop diagrams, and short-range interactions.
The lowest order, where four--nucleon contact interactions start to contribute,
is N$^4$LO. To estimate their strength,
Ref.~\cite{Gardestig}
evaluated certain tree-level contributions as indicated by
Figs.~\ref{ddlettdiag}c.
This figure represents the exchange of heavy mesons
($\sigma$, $\omega$, $\rho$) via a Z-graph mechanism,
with $\pi$-$\eta$ mixing to generate CSB at pion emission
(${\cal M}_\sigma$, ${\cal M}_\omega$, and ${\cal M}_\rho$).
Another Z-graph (labeled as ${\cal M}_{\rho\omega}$) 
arises in which the CSB occurs  in the heavy-meson exchange via 
 $\rho$-$\omega$ mixing along with strong  pion emission at the vertex.

The Z-graphs are believed to be important, because their inclusion lead to a
quantitative description of the total cross section for the reaction $pp\to
pp\pi^0$ near threshold 
\cite{lowe}. The  results \cite{csbnogga} use the coupling constants  and parameters of
Ref.~\cite{Gardestig}, see their Table~I. 
It was  found that  the Z-graphs give unexpectedly large
contributions, especially the $\rho$-$\omega$ exchange operator that add constructively
and  overwhelm
the one-body term. This model of  resonance saturation, gives 
 results in vast disagreement with the power counting and therefore needs reassessment.
 Here we re-asses
 the coupling constants used by \cite{csbnogga}.  This is only a first step,
because it is also necessary to justify the use of Z-graphs in resonance saturation, which would require further calculations.

\vspace{.2cm}  The various mechanisms generate pion-production kernels
that are  sandwiched between  final and initial state wave functions  to provide a
transition matrix element ${\cal M}$ for  $T_d$=228.5 MeV.  These matrix elements are given in
Table~1 of \cite{csbnogga}.
The transition amplitude can be written as
\bea {\cal M}&=&{\cal M}_{\rm PE}+{\cal M}_{\rm rec} +{\cal M}_{\rm 1b}+{\cal M}_{\sigma}+{\cal M}_{\rho}+{\cal M}_{\omega} 
\nonumber \\ &&
+{\cal M}_{\rho\omega}+{\cal M}_{\rm WF},\eea
where the pion exchange term ${\cal M}_{\rm PE}$, its recoil correction ${\cal M}_{\rm rec}$,
 and the effects of CSB in the $\alpha $ wave function ${\cal M}_{\rm WF} $ are independent of 
$g_\eta$ and $\beta_1$. 
The one-body term and sigma and rho exchange terms involving the Z-graphs, 
${\cal M}_{\rm 1b}+{\cal M}_{\sigma}+{\cal M}_{\rho}$,
 are proportional to $\beta_1$. 
The terms $ {\cal M}_{\omega} +{\cal M}_{\rho\omega}$,
which arise from omega and rho-omega exchange, 
are proportional to $\beta_1$ and $g_\omega$, the strong $\omega$-nucleon coupling constant.
Given the contributions to ${\cal M}$ expressed in Table~1 of \cite{csbnogga} 
in units of $10^{-4}\;{\rm fm}^{-2}$,
 the cross section can be written as
\bea \sigma=4.303 \;{\rm pb}\left |{\cal M}\;\times 10^4 \;{\rm fm}^2\right|^2.\label{cross}\eea

Results were obtained  \cite{csbnogga} using
either the Argonne $V_{18}$ (AV18) \cite{Wiringa:1994wb} 
or CD-Bonn (CDB) \cite{Machleidt:2000ge} 
two-nucleon potentials  combined with a properly adjusted 
Tucson-Melbourne (TM99) \cite{tm99} three-nucleon force. 
The combination guarantees
that the $\alpha$ particle binding energy is 
reproduced with high accuracy.

We now turn to CSB in the nucleon-nucleon (NN) system due to meson-mixing. 
Our CSB NN calculation is described in Ref.~\cite{MM01}.
Since we have to use meson-nucleon form factors in our NN scattering
calculations, while in Refs.~\cite{Gardestig,csbnogga}
no such form factors are applied,
we explain the definition 
of meson-nucleon coupling constants in conjunction with form factors.
For this, we define a coupling constant as a function of the momentum-transfer
$t$ by
\begin{equation}
g_\alpha(t)\equiv F_\alpha(t) g_\alpha (t=m_\alpha^2)
\end{equation}
with
\begin{equation}
F_\alpha(t) = \frac{\Lambda_\alpha^2-m_\alpha^2}{\Lambda_\alpha^2-t} \,,
\end{equation}
where $\alpha$ stands for any meson, $m_\alpha$ denotes the meson mass
and $\Lambda_\alpha$ the so-called cutoff mass. 
We use $\Lambda_{\rho,\omega}=1400$ MeV.
In the relativistic three-dimensional theory used for CD-Bonn,
the momentum transferred between the two nucleons is
$t=-({\bf q'} - {\bf q})^2$ with $\bf q$ and $\bf q'$ the center-of-mass
nucleon three-momenta before and after scattering.
Coupling constants used in theories without form factors should be
compared to coupling constants at $t=0$, i.e., $g_\alpha(0)$,
of a theory with form factors. Thus, in our CSB NN calculations, we use
the coupling constants 
given in Table~I of Ref.~\cite{Gardestig}
and identify them with $g_\alpha(0)$, except for the omega-nucleon
coupling where we use
\begin{equation}
\frac{g^2_\omega(0)}{4\pi} = 5.0
\label{eq_omega}
\end{equation}
instead of the 10.6 applied in Ref.~\cite{Gardestig}.
Our lower value for the omega coupling is more consistent with
a fixed-$s$ dispersion relations analysis 
by Hamilton and Oades~\cite{HO84} in which
a value of $5.7\pm2.0$ was obtained. Moreover,
$SU(3)_{\rm flavor}$ implies $g_\omega=3 g_\rho$
which, for the rho-coupling used, yields $g^2_\omega/4\pi=3.9$.
Applying the omega coupling constant stated in Eq.~(\ref{eq_omega}) and
the other meson parameters
as given in Table~I of Ref.~\cite{Gardestig}
and using
$\langle\rho^0|H|\omega\rangle=-4300$~MeV$^2$
for the $\rho$-$\omega$--mixing matrix
element~\cite{mesmix}, 
we find a CSB contribution to the $^1S_0$
scattering length difference from $\rho-\omega$ mixing of
\bea \Delta a_{\rho\omega}=1.45\;{\rm fm}\,.\eea
Obviously, this term entirely accounts for the observed 
CSB scattering length difference. We have also calculated
 $\eta-\pi^0$ mixing to find
\bea \Delta a_{\pi\eta}=0.33 \;{\rm fm}\eea 
using the same strong coupling constants and mixing matrix element as in
\cite{csbnogga}. Since the $\rho\omega$ mixing  explains all CSB, 
this might be considered as yet another argument that $g_\eta=0$.

To explore a variety of possibilities, we re-write the results of \cite{csbnogga} as
\bea 
{\cal M}_{\rm CDB} & = & \left(-2.75 + 3.1 i + (-3.4 + 2.82 i) 
\sqrt{\frac{g_\eta^2/4\pi}{.51}} 
\right. \nonumber \\ && \left.
+ (-.53 + .44 i) \sqrt{\frac{g_\eta^2/4\pi}{.51}}\,\frac{g_\omega^2/4\pi}{10.6} 
\right. \nonumber \\ && \left.
   + (-1.32 + 1.51 i) \sqrt{\frac{g_\omega^2/4\pi}{10.6}} 
\right. \nonumber \\ && \left.
+ .51 - .13 i\right)10^{-4}\;{\rm fm}^{-2},
\eea
or 
\bea
 {\cal M}_{\rm AV18} & = & \left(-1.65 + 1.58 i + (-2.44 + 2.21 i) 
\sqrt{\frac{g_\eta^2/4\pi}{.51}} 
\right. \nonumber \\ && \left.
+ (-.35 + .34 i) \sqrt{\frac{g_\eta^2/4\pi}{.51}}\,\frac{g_\omega^2/4\pi}{10.6} 
\right. \nonumber \\ && \left.
   + (-0.84 + 1.07 i) \sqrt{\frac{g_\omega^2/4\pi}{10.6}} 
\right. \nonumber \\ && \left.
+ .41 - .14 i\right)10^{-4}\;{\rm fm}^{-2},
\eea
depending on the potential used. 
Ref.~\cite{csbnogga} observed that this
model dependence  visibly influences the   cross section result, requiring 
a more consistent treatment of the  NN interaction and 
production operator in the future.

 As a first step, we note that using the parameters of \cite{csbnogga} leads to the results
 \bea{\cal M}_{\rm CDB}&=&(-7.49 + 7.74 i)10^{-4}\;{\rm fm}^{-2},
\\
{\cal M}_{\rm AV18}&=&(-4.87 + 5.06i)10^{-4}\;{\rm fm}^{-2},\eea
with cross sections
\bea
\sigma_{\rm CDB} &=& 499 \; {\rm pb},
\\
\sigma_{\rm AV18} &=& 212 \; {\rm pb}.
\eea

Next we replace $g_\omega^2/4\pi$ by the value stated in Eq.~(\ref{eq_omega}),
which yields
  \bea{\cal M}_{\rm CDB}&=&(-6.80 + 7.03 i)10^{-4}\;{\rm fm}^{-2},
\\
{\cal M}_{\rm AV18}&=&(-4.42 + 4.54i )10^{-4}\;{\rm fm}^{-2}.\eea
  This would lead to about a 20\% reduction of the cross section.
  However, using $\pi\eta$ mixing along with $\rho\omega$ mixing overestimates $\Delta a$. 
Therefore, it is reasonable to explore the consequences of using $g_\eta=0$,
which results in
  \bea{\cal M}_{\rm CDB}&=&(-3.15 + 4.00 i)10^{-4}\;{\rm fm}^{-2},
\\
{\cal M}_{\rm AV18}&=&(-1.82 + 2.17i )10^{-4}\;{\rm fm}^{-2},\eea  
with cross sections
\bea
\sigma_{\rm CDB} &=& 111.5 \; {\rm pb},
\\
\sigma_{\rm AV18} &=& 34.5 \; {\rm pb}.
\eea
The cross section for AV18 is now only a factor of 2 or so bigger than the data, 
while CD-Bonn is off by a factor of about 7.

In summary, past theoretical work~\cite{csbnogga}
 on the cross section of the
$dd\rightarrow\alpha\pi^0$ reaction 
at 228.5 MeV was plagued
by the problem that 
the predictions were off
by factors between 15 and 30.
In this note, we have shown that constraining the coupling constants
involved by the requirement that the CSB in the $^1S_0$ NN scattering length
is correctly reproduced reduces the over-prediction to just a factor of about 2
(using the AV18 potential). In relative terms, this is substantial
progress in understanding the
$dd\rightarrow\alpha\pi^0$ reaction. However, significant differences
 between the use of the AV18 and CD-Bonn potentials remains,
 signaling that a deeper understanding is needed.

\section*{Acknowledgments}

We thank C.~Hanhart 
for useful discussions and encouragement. This research was partially funded by
FCT grant POCTI/37280/FNU/2001 (ACF) and
DOE grants DE-FG02-97ER41014 (GAM) and
DE-FG02-03ER41270 (RM).

\bibliographystyle{unsrt}

\end{document}